\documentclass[pre,twocolumn,showpacs,preprintnumbers,amsmath,amssymb]{revtex4}
\usepackage{epsfig}
\usepackage{graphicx}
\usepackage{longtable}
\usepackage{dcolumn}
\usepackage{amsmath}
\usepackage{amssymb}
\usepackage{bm}
\usepackage{amsmath}
\begin{document}

\title{Excitable systems with noise and delay with applications to control: renewal theory approach}

\author{Andrey Pototsky} 
\author{Natalia Janson}
\affiliation{Department of Mathematical Sciences, Loughborough University,
  Loughborough, Leicestershire LE11 3TU, United Kingdom}

\date{\today}

\begin{abstract}
We present an approach for the analytical treatment of excitable systems with
noise-induced dynamics in the presence of time delay. An excitable system is modeled 
as a bistable system with a time delay, while another delay enters as a control term taken after [Pyragas 1992] as a difference between the current system state and its state $\tau$ time units before. 
This approach combines the elements
of renewal theory  to estimate the essential features of the resulting stochastic process as functions of the
parameters of the controlling term. 
\end{abstract}


\maketitle

\section{Introduction}
\label{intro}

Oscillations induced by noise are observed in many nonlinear systems with dissipation including neural networks,
lasers,
chemical reactions, 
biomembranes, etc. (see \cite{Lind04} and references therein). 
At present, three different types of systems  are being recognized 
that demonstrate noise-induced oscillations: bistable systems \cite{Lind00}, systems close to Andronov-Hopf bifurcation \cite{Nei97}, and excitable systems \cite{Lind04}.


Much effort has been put into the development of the analytic description of excitable systems. 
The approach of the renewal theory \cite{Cox} was successfully adopted to describe the real excitable systems with continuous dynamics by a two-state \cite{Lind00}, or by a three-state \cite{Goy05,Prag03} discontinuous stochastic process. In the frames of the renewal theory, it is assumed that the system is allowed to be only in a finite number of states, and the noise makes 
the system switch between the states in a random manner. 
 The time the system spends in a given state before undergoing the next transition is called residence  time in this state. For the validity of the renewal theory it is crucial that 
 the distribution function of the residence time in a given state does not change in time.
Therefore, in the case of an $n$-state system, its entire dynamics is determined by $n$ distribution densities of the residence times, also known as residence time densities (RTDs). For the FitzHugh-Nagumo system the RTDs were explicitly calculated in \cite{Lind00} using the Fokker-Planck equation approach suggested in \cite{Mel93}. Phenomenological three-state model of the excitable system with the arbitrary RTDs was discussed in \cite{Goy05,Prag03}.

The situation becomes more complicated when a delay term is introduced into an excitable system, and the random process in it becomes non-Markovian. This problem arises for example in relation to the problem of controlling noise-induced motion. 

Usually noise-induced oscillations possess a certain  timescale  that is dependent on the parameters of the applied noise, e.g. its intensity. One possible way to define this timescale for an excitable system 
is to estimate the power spectrum that would normally contain one or more peaks, and to take the inverse of the frequency of the highest peak to be the main period of oscillations. In \cite{Janson04} the idea was introduced to control the properties of oscillations induced merely by external noise by applying time-delayed feedback force $F(t)$ in the Pyragas form \cite{Pyragas92}: $F(t)=k(x(t-\tau)-x(t))$, where $\tau$ is time delay and $k$ is the feedback strength. It has been shown that the timescale and the coherence of noise-induced oscillations can be changed by adjusting solely the delay time in the controlling force. Moreover, an almost piecewise-linear dependence of the main period on the delay time was revealed. Remarkably, a similar behaviour of the main period was found in systems with time-delayed feedback that were either excitable, or close to Andronov-Hopf bifurcation \cite{Janson04,Balanov04}. For a van der Pol oscillator this phenomenon was explained  by means of computation of the power spectrum of oscillations analytically: by using either linearization  \cite{Schoell05,Pototsky07,Jan07}, or mean-filed approximation \cite{Amann05,Jan07}. 

The theory of the excitable systems with time delay is still missing. Some progress was made in  \cite{PSLS07}, where the impact of the delayed feedback on the current dynamics of the system was considered in the mean field approximation. Here we propose an alternative approach based on the analysis of renewal processes with history-dependent RTDs. As mentioned above, the renewal theory can only be applied in the case when the RTDs {\it do not change in time}. This condition is obviously violated for the  processes with time delay, in which RTDs are dependent on history. 

In order to overcome this problem, we introduce {\it equilibrium} RTDs by averaging over all possible histories. This approach is somewhat similar to the one suggested in \cite{McNam88} for calculating the RTD in the case of stochastic resonance, where the noise-induced switchings between the two potential wells are considered  in the presence of a weak periodic perturbation. Namely, in \cite{McNam88} the unknown RTD is obtained by averaging the known  distribution of escape times (time it takes to escape from a certain well) over the known probability that the system spends a certain amount of time in the left well before entering the right well. In contrast to this, we show that in the case of {\it history-dependent} renewal process, the equilibrium RTDs are given by the solution of an integral equation. This equation is derived for an arbitrary, and solved analytically for a moderate, delay time.

In order to compare the analytic results of the modified renewal theory 
with the numerical results for an excitable system, we design an excitable system from a bistable system with a non-symmetric potential and with a time delay, following the idea of the delay-induced excitability proposed in \cite{Piw05}.
To this system we add the controlling force $F(t)$  as above. 
The parameters of the two-state model used for analytics are matched to the parameters of the bistable  system via the Kramers formula for the transition rates \cite{Kra40,Kram,Gard}. We employ 
the hybrid approach by combining the renewal theory results with the equilibrium RTDs of the history-dependent process in order to approximate the power spectrum of the noise-induced oscillations.  
Incoherence maximization due to delay is demonstrated for positive feedback strength $k$ only.
%
\section{The model}
\label{sec1}

\subsection{Excitability}
\begin{figure}
\begin{center}
\includegraphics[width=0.45\textwidth]{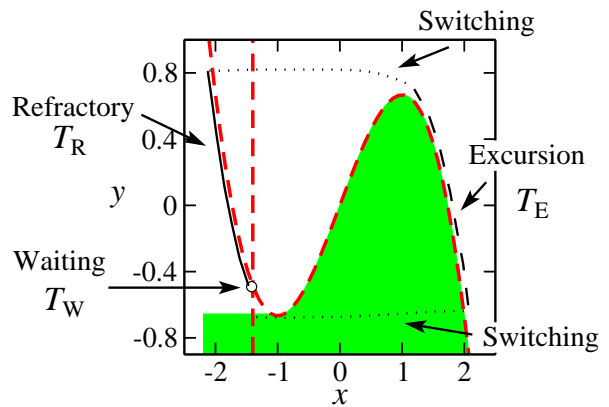}
\caption{\label{fig0} Phase plane of FitzHugh-Nagumo system Eqs.~(\ref{fhn}) in an excitable regime.
Dashed grey lines -- null-clines, empty circle -- fixed point, shaded area -- the area from which an excursion can start. Stages of one full oscillation are shown: black dashed line -- excursion, black solid line -- refractory, dotted lines -- switchings between the two branches (very fast). }
\end{center}
\end{figure}

\begin{figure}
\begin{center}
\includegraphics[width=0.45\textwidth]{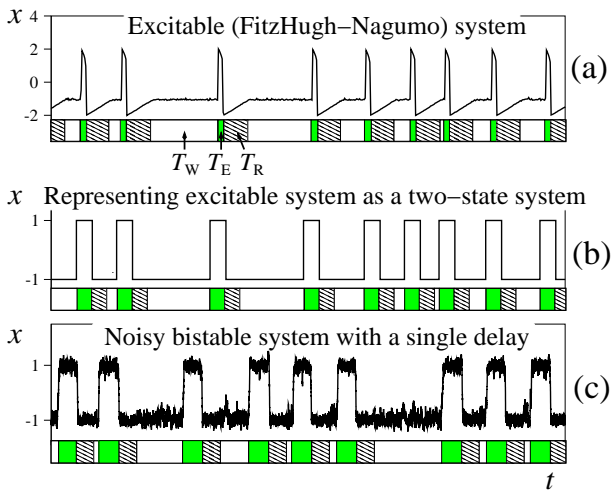}
\caption{\label{fig1} (a)  Realization $x(t)$ of an excitable system Eq.~(\ref{fhn}); (b) 
Stages of oscillations in (a) modeled by a two-state process; (c) Realization of a noisy bistable system with delay Eq.~(\ref{1eq2}). \\
Lower panels of (a)-(c) show stages of the process: white -- waiting with duration $T_W$, shaded -- excursion with duration $T_E$, patterned -- refractory with duration $T_R$.  }
\end{center}
\end{figure}

Before describing the way to construct an excitable system from a bistable system with delay following \cite{Piw05}, we need to explain the concept of excitability. 
A popular toy model of an excitable system is a FitzHugh-Nagumo system
\begin{eqnarray}
 \dot{x}&=&\bigg( x-\frac{x^3}{3}-y  \bigg) \frac{1}{\epsilon}, \nonumber \\
\dot{y}&=& x+a+D \zeta(t), \label{fhn}
\end{eqnarray}
where $\zeta(t)$ describes random fluctuations with Gaussian distribution, zero mean and unity variance. 
The null-clines of this system, which are the curves defined by $\dot{x}$$=$$0$ and $\dot{y}$$=$$0$
(assuming that $D$$=$$0$) are shown in Fig.~\ref{fig0} by grey dashed lines. Usually this system is considered at $\epsilon$$ \ll$$ 1$ in order to provide timescales separation described below. 
At $a$$>$$1$ the system has a single stable fixed point (empty circle in Fig.~\ref{fig0}), and no oscillatory dynamics without noise ($D$$=$$0$). When Gaussian noise is applied ($D$$>$$0$), the behavior of the system changes drastically. Namely, while the value of  $D \xi(t)$ is small, the system oscillates randomly around the fixed point during the waiting phase of duration $T_W$ (note the horizontal plateaus of the realization of $x(t)$ in Fig.~\ref{fig1}(a)). But when the values of $D \xi(t)$ are large enough
to throw the phase point into grey area in Fig.~\ref{fig0}, the system quickly tends to the right-hand branch of the cubic parabola along an almost horizontal path (lower dotted line in Fig.~\ref{fig0}), since due to the smallness of $\epsilon$ the horizontal component $\dot{x}$ of the phase velocity is much larger than its vertical component $\dot{y}$. 
Then the system enters its excursion stage in which the phase point slowly crawls upwards along 
the right-hand branch of the parabola during $T_{E}$ time units (black dashed line in Fig.~\ref{fig0}): this motion is smeared by noise,
but its velocity is almost unaffected by the latter, at least in average. 
As soon as the phase point reaches the top of the right-hand parabola branch, the vector flow swiftly carries it towards the left-hand branch (upper dotted line in Fig.~\ref{fig0}). Finally, the system enters its refractory stage as the phase point
slowly crawls downwards towards the fixed point during time $T_R$ (black solid line in Fig.~\ref{fig0}), again almost unaffected by noise in average. Then the process is repeated.  

A typical realization of a stochastic process $x(t)$ occurring in  system Eqs.~(\ref{fhn}) looks similar to the profile shown in Fig.~\ref{fig1}(a), where an $x$-variable from the FitzHugh-Nagumo system in the excitable regime is plotted. The cells in the lower panel show different stages of the process: white -- waiting, shaded -- excursion, patterned  -- refractory. 
One can single out two essential features of this motion induced merely by external noise. First, one can distinguish between very fast motion between the two branches of the cubic parabola, and slow motion 
along the branches. If $\epsilon$ is very small, one can assume that the switching between the branches occurs instantly, i.e. the system can be  only in one of the two states corresponding to the two
branches of the parabola. Second, the durations $T_E$ and $T_R$ of the excursion and of the refractory stages can be approximately regarded as independent of noise and constant in any event of a large excursion in the phase space, while the duration $T_W$ of the waiting stage is completely determined by noise. The distribution density of $T_W$ depends both on noise and on the design of the system. 

The dynamics of the FitzHugh-Nagumo system in the excitable regime can be approximated by a two-state stochastic process $s(t)$ which can take only two values, say, $s= \pm 1$ (Fig.~\ref{fig1}(b)). 
We assume that in the excursion stage the system is in state $s=+1$, and in the waiting stage in  $s=-1$. In addition, 
we assume that in the refractory stage $s(t)$ does not change and assign $s=-1$ like in the waiting stage. 
Therefore, residence time in the state $(-1)$ is equal to the sum of the refractory and waiting times, $(T_R+T_W)$.

The dependence of the transition rate $\lambda$ on the parameters of the FitzHugh-Nagumo system is rather complex and can be obtained only numerically. However,  for a bistable system this information is readily available  via the Kramers formula \cite{Kra40,Kram,Gard}.
One can also neglect the intra-well dynamics of a bistable system and match the parameters of the latter with the parameters of a two-state system. 

\subsection{Bistable excitable system with delayed feedback}

The idea of the present approach is to construct an excitable system from a bistable system by including a delay term into the latter, following the concept of delay-induced excitability proposed in \cite{Piw05}.  
Namely, in \cite{Piw05} it was shown that a two-state system with {\it a single} time delayed feedback $T$ can behave as an excitable one if the double-well potential describing the system is asymmetric. A characteristic feature of this model is that  there is a locking of state, i.e. the transition from one state into another is allowed to occur not earlier than $T$ seconds after the moment of the previous transition.   This feature is illustrated in Fig.~\ref{fig1}(c) where a realization of such a bistable system is shown, with the deeper potential well being located around the value of $x=-1$. 
Here, one can recognize the same stages that occur in an excitable system (compare with (a)): waiting marked as a white cell in lower panel, excursion marked by shaded, and refractory marked by a patterned cell. Note, that in the bistable system with a single delay $T$ the following condition is automatically satisfied by the way of construction:  $T_E = T_R=T$.  
As a result, the distribution densities of the waiting times in both states  are shifted to the right by $T$ as compared to the case without delay. 

We would like to assess the effect of the controlling term $F(t)$ on this system, in which the second delay appears. 
The second delay term is due to the delayed feedback force.

The dynamics of the bistable system with two time delays is described by the equation
\begin{eqnarray}
\dot{x}(t) =  -\frac{\partial U[x,x_{T} ]}{\partial x} + k  [x_{\tau}-x] + \sqrt{2 D} \zeta(t),
\label{1eq2}
\end{eqnarray}
where $x_{T}$ and $x_{\tau}$ denote the retarded variables $x(t-T)$ and $x(t-\tau)$, respectively; $T$ is the fixed excursion and refractory time, $\tau$ is the delay time of the controlling force, $k$ is the strength of the controlling force, $\zeta(t)$ is a Gaussian noise, and $D$ is the noise strength. 

The potential $U[x,x_{T}]$ in Eq.~(\ref{1eq2}) is chosen to have two wells with minima located at $x = \pm 1$ that are separated by a barrier with the maximum at $x=-0.3$, so that
\begin{eqnarray}
U[x,x_{T}] &=& \frac{x^4}{4} -\frac{x^2}{2} - (0.1+0.2 x_{T}) (x^3 -3 x), \label{1eq3}\\
\frac{\partial U[x,x_{T}]}{\partial x}
&=&(x^2-1) (x-0.3-0.6 x_{T} ).  \nonumber
\end{eqnarray}
Following \cite{Kra40}, we assume the overdamped case when we can neglect the intra-well dynamics. It is well known that this approximation is valid for noise intensities much smaller than the height of the potential barrier, i.e. for $D \ll \Delta U_{\rm eff}$.

The special choice of the position of two minima does not affect the results presented below. 

The evolution equation Eq.~(\ref{1eq2}) can also be rewritten in terms of an effective potential $U_{\rm eff}$ which includes the controlling force
\begin{eqnarray}
U_{\rm eff}[x,x_{T},x_{\tau}] = U[x,x_{T}] + k\frac{x^2}{2}-kx_{\tau}x.
\label{1eq4}
\end{eqnarray}
We will be considering small values of feedback strength  $k$ for which $x_{T}$ and $x_{\tau}$ 
can be either in the left or in the right well. With this, the values they are taking are close to $-1$ or to $+1$, respectively, therefore we single out two states of the system that we denote as $(-1)$ and $(+1)$. 
In the analytical calculations below we will substitute the exact instantaneous values of $x_{T}$ and $x_{\tau}$  by
their approximate values $\pm 1$. 

\begin{figure}
\begin{center}
\includegraphics[width=0.45\textwidth]{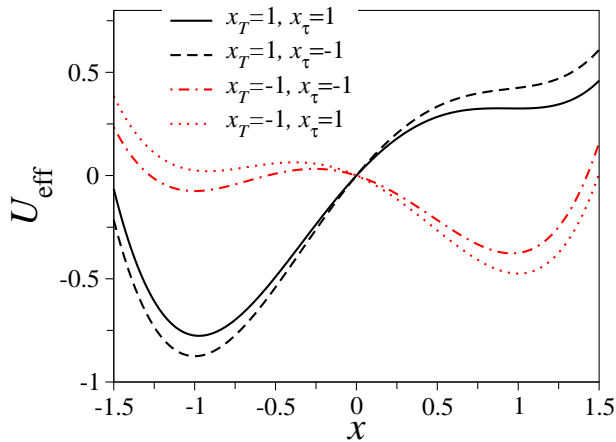}
\caption{\label{fig2} Effective potential $U_{\rm eff}[x,x_{T},x_{\tau}]$ (\ref{1eq4}) for $k=0.05$ and fixed $x_{T}$ and $x_{\tau}$ as in the legend.}
\end{center}
\end{figure}

In Fig.~\ref{fig2} the effective potential $U_{\rm eff}$ is shown for fixed $x_{T}$ and $x_{\tau}$ that take values $\pm 1$. Consider all stages of one oscillatory cycle. Start from $x_{T}\approx+1$.  From  Fig.~\ref{fig2} it is clear that at any $x_{\tau}$ there exists only one well (left) and  the particle trapped in it will remain there until $x_{T}$ switches  to approximately $(-1)$. This is the refractory phase with duration $T_R$. At $x_{T}$$\approx$$-1$ the right well appears and  the waiting phase begins, during which it is possible to jump from the left to the right well. Transition rate $\lambda$ from $(-1)$ to  $(+1)$ becomes dependent on $x_{\tau}$: for $k>0$ it is larger when $x_{\tau} \approx +1$, i.e. when the left well is shallower. After the particle jumps into the right well,  the system is in the excursion phase. Note that at $x_{T} \approx -1$ the right minimum is much deeper than the left one. Therefore, the particle trapped in it will remain there until $x_{T}$ changes from $(-1)$ to $(+1)$, i.e. until the right well vanishes.  After that the particle jumps  from state $(+1)$ to the state $(-1)$ and the cycle repeats again.

\begin{center}
\begin{table}[h]
\label{tabs}
\begin{tabular}{|c|c|c|c|c|}
\hline
& \begin{tabular}{c}$x_{T}=+1$\\$x_{\tau}=+1 $\end{tabular} &\begin{tabular}{c}$x_{T}=+1$\\$x_{\tau}=-1 $\end{tabular} &\begin{tabular}{c}$x_{T}=-1$\\$x_{\tau}=+1 $\end{tabular} & \begin{tabular}{c}$x_{T}=-1$\\$x_{\tau}=-1 $\end{tabular} \\ 
\hline 
$(-1)\rightarrow (+1)$ & $0$ &$0$ &$p +q$  &$p$\\
\hline
$(+1)\rightarrow (-1)$ & $\infty$ & $\infty$ & $0$ & $0$\\
\hline
 \end{tabular}
\caption{ \label{tab1} Transition rates $\lambda$ between the states of the bistable system with delay
at different values of $x_{T}$ and $x_{\tau}$.}
 \end{table}
\end{center}
The information on the transition rates is summarized in Table~\ref{tab1}. Here we denote by ($p + q$), $q$\,$>$\,$-p$, the transition rate from the state $(-1)$ to the state $(+1)$ if $x_{\tau}$\,$\approx$\,$+1$ and $x_{T}$\,$\approx$\,$-1$ and by $p$, $p$\,$>$\,$0$, the the transition rate from the state $(-1)$ to the state $(+1)$ if $x_{\tau}$\,$\approx$\,$-1$ and $x_{T}$\,$\approx$\,$-1$.
The transition rates are calculated from $U_{\rm eff}$ according to the Kramers theory \cite{Kra40} as follows

\begin{eqnarray}
p &=& \frac{1}{2 \pi}\sqrt{-\partial_{xx}U_{\rm eff} (x_m,-1,-1)\partial_{xx}U_{\rm eff} (x_0,-1,-1)}\nonumber \\ 
&&\exp{\left[-\frac{\Delta U_{\rm eff}}{D}\right]} \nonumber \\
p +q &=& \frac{1}{2 \pi}\sqrt{-\partial_{xx}U_{\rm eff} (x_m^{\prime},-1,1)\partial_{xx}U_{\rm eff} (x_0^{\prime},-1,1)}\nonumber \\
&&\exp{\left[-\frac{\Delta U_{\rm eff}'}{D}\right]},
\label{1eq5}
\end{eqnarray}
where $x_0$ and $x_m$ are the positions of the maxima and the left minima of the effective potential $U_{\rm eff}$ at $x_{T} = x_{\tau} = -1$ and $x_0^{\prime}$ and $x_m^{\prime}$ are the same quantities at $x_{T} =-1$, $x_{\tau} = +1$. $\Delta U_{\rm eff}$ and $\Delta U_{\rm eff}'$ are the potential differences between the maxima and the minima. Note, that due to presence of the feedback term $k(x_{\tau}-x)$, the values of $x_m$ and $x_m'$ are not exactly equal to $-1$, and $x_0$ is not exactly $-0.3$. However, in analytic calculations we substitute $x_0 = x_0^{\prime}$,  $x_m$ and $x_m'$ by their approximate values $-0.3$, $-1$ and $-1$, respectively.  Throughout the paper we compare the analytic results derived from the two-state model with the results of simulation of a bistable system with two time delays.

\section{Two-state model: analytic results}
\label{sec2}
One should notice that the two-state model is not derived from the continuous bistable system Eq.\,(\ref{1eq2}) with linear feedback term. 
It can describe a larger class of systems with similar properties, e.g.  systems with  non-linear feedback term, provided that the transition rates depend on the delay time according to the phenomenological rule provided by Table\,(\ref{tab1}).

 The continuous random process $x(t)$ is approximated by the discrete random process $s(t)=\pm 1$ with infinitely fast (discontinuous) transitions from one state to the other. In order to match the parameters of the two-state model with those of the bistable system Eq.~(\ref{1eq2}), the excursion and the refractory times must be set equal, i.e. $T_E$ $=$ $T_R$ \cite{Piw05}. However for the sake of generality the analytic results presented below were obtained for the case of $T_E \not= T_R$.
In what follows by $\xi$ we denote the time of residence in a certain state: from the context it will be clear what state or phase is referred to. 

Regarding the dependence of the transition rates on the history (Table~\ref{tab1}) we conclude that the RTD $\psi_{+}(\xi)$ in the state $s=+1$  is given by the delta-function $\psi_{+}(\xi) = \delta(\xi - T_E)$. This assumption is justified if there is a strong time scale separation meaning that the transition from the excited to the non-excited state and backwards occurs almost instantaneously. Moreover, the RTD in the state $s=-1$ is zero for the first $T_R$ seconds after the transition from the state $s=+1$ to the state $s=-1$. Consequently, the model contains only one unknown object, namely the distribution density  $\psi_{-}(\xi)$ of the waiting time. It is related to the RTD $\Psi_{-}(\xi)$ in the state $s=-1$, via
\begin{eqnarray}
\Psi_{-}(\xi) = 
\left\{
\begin{array}{cl}
0,&\xi \in [0;T_R]\\
\psi_{-}(\xi-T_R), & \xi \in [T_R;\infty)
\end{array}
\right.
\label{RTD1}
\end{eqnarray}
\subsection{Small delay times $\tau$: non-variable history}
\label{sec2a} 
\begin{figure}
\begin{center}
\includegraphics[width=0.45\textwidth]{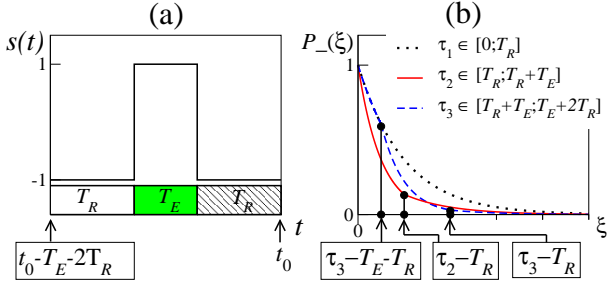}     
\caption{\label{fig2a} (a) Sketch of the profile of the two-state process $s(t)=\pm 1$ on the interval of time $[t_0-T_E - 2 T_R;t_0]$. (b) Three possible forms of the survival probability $P_{-}(\xi)$ in the waiting phase  for $\tau < T_E + 2 T_R$. Here, $\xi$ is the duration of the waiting phase. $P_{-}(\xi)$  is shown for three different values $\tau_i$ of the delay time $\tau$ chosen arbitrarily from the intervals indicated in the legend. Filled circles on the $\xi$-axis indicate points where the change of the slope of the corresponding $P_{-}(\xi)$ occurs.}
\end{center}
\end{figure}
We start with the case of small delay times from the interval $\tau \in [0;T_E+2 T_R]$. 
The  probability $P_{-}(\xi)$ of survival in the state $(-1)$ during the waiting phase is given by the solution of the equation
\begin{eqnarray}
-\psi_{-}(\xi)=\frac{\partial P_{-}(\xi)}{\partial \xi} = - \lambda_{-}(\xi) P_{-}(\xi),
\label{2eq1}
\end{eqnarray}
where $\lambda_{-}(\xi)$ stands for the transition rate from  $s=-1$ to $s=+1$. 
If $\lambda$$=$${\rm const}$, which is valid for the bistable systems without delays, the statistics of the waiting times is exponential \cite{Cox} with the distribution density $\psi$ given by
\begin{equation}
\psi (\xi) = \lambda \exp{(-\lambda \xi)},
\label{1eq1}
\end{equation}
 where the constant transition rate $\lambda$ depends on the system parameters. The mean waiting time $\langle T_W \rangle$ is then given by $\langle T_W \rangle = 1/ \lambda$.
However, due to delay there appears a discontinuity in the transition rate $\lambda_{-}(\xi)$ (see Table 1), therefore the solution of the Eq.~(\ref{2eq1}) on the intervals where $\lambda_{-}(\xi)$ is constant must be normalized in such a way that the survival probability $P_{-}(\xi)$ remains continuous. 

Suppose that at time $t=t_0$  (Fig.~\ref{fig2a}(a)) the refractory phase has just finished and the system is in the very beginning of the waiting phase, implying that the waiting duration is equal to  zero, $\xi=0$. Fig.~\ref{fig2a}(a) shows the profile of the two-state process on the interval of time  $t \in [t_0-T_E - 2 T_R;t_0]$.
It is clear that no other profile $s(t)$ is possible on this interval of time. This means that despite the fact that the solution of Eq.~(\ref{2eq1}) depends on $\tau$, this dependence remains tha same during any waiting phase of any cycle, as long as $\tau$ is less than $T_E+2T_R$.
Hence, the distribution density of the residence times in the state $s$\,$=$\,$-1$ does not change in time and the renewal theory \cite{Cox,Goy04} can be applied. 
Three different cases should be considered separately for $\tau \le T_E + 2 T_R$.

{\it Case 1:} $\tau \in [0;  T_R]$. The solution of Eq.~(\ref{2eq1}) is then given by
\begin{eqnarray}
P_{-}(\xi) = 
\begin{array}{l}
e^{-p \xi},~~~\xi \in [0;\infty).
\end{array}
\label{2eq2}
\end{eqnarray}
This solution is shown by the dotted line in Fig.~\ref{fig2a}(b).
 
{\it Case 2:} $\tau \in [T_R;  T_E + T_R]$. In this case the survival probability reads
\begin{eqnarray}
P_{-}(\xi) = 
\left\{
\begin{array}{lcl}
e^{-(p+q) \xi},&\ & \xi \in [0;\tau-T_R]\\
e^{-p \xi}e^{-q (\tau -T_R)},&\ &\xi \in [\tau-T_R;\infty).
\end{array}
\right.
\label{2eq3}
\end{eqnarray}
This solution is shown by the solid line in Fig.~\ref{fig2a}(b).

{\it Case 3: $\tau \in [T_E + T_R; T_E + 2 T_R]$.} 
\begin{eqnarray}
P_{-}(\xi) &=& e^{-p\xi} \label{2eq4} \\
&\times&\left\{
\begin{array}{lcl}
1,&\ &\xi \in [0;\tau-T_E - T_R]\\
e^{-q  \xi}e^{q( \tau-T_E - T_R) },&\ &\xi \in [\tau-T_E - T_R;\\
&\ & \qquad \tau-T_R]\\
e^{-q T_E },&\ &\xi \in [\tau-T_R;\infty).
\end{array}
\right.\nonumber 
\end{eqnarray}
Dashed line in Fig.~\ref{fig2a}(b) corresponds to $P_{-}(\xi)$ in the {\it Case 3}.

We can define the main period $T_{\rm main}$ as $T_{\rm main} = 2 \pi / \Omega_{\rm max}$, where the frequency $\Omega_{\rm max}$ corresponds to the absolute maximum of the power spectrum of oscillations.   

In \cite{Janson04} it has been shown numerically for FitzHugh-Nagumo system that $T_{\rm main}$ depends almost piecewise-linearly on $\tau$.
It was assumed there that the piece-wise linearity in the main period {\it vs} delay time is a universal phenomenon that should occur in any system driven by noise including non-excitable systems near Andronov-Hopf bifurcation. 
Here, a similar dependence of $T_{\rm main}$ on $\tau$ will be demonstrated for system Eq.~(\ref{1eq2}) analytically already for $\tau \in [0;T_E + 2 T_R]$ and $k$$<$$0$.
From Eqs.~(\ref{2eq2})--(\ref{2eq4}) one can calculate the power spectrum of oscillations using the result of the renewal theory \cite{Cox,Goy04}
\begin{eqnarray}
S(\omega) = \frac{2(\Delta x)^2}{\langle T_{+} \rangle + \langle T_{-} \rangle}\frac{1}{\omega^2}Re\left\{ \frac{[1-\tilde{\Psi}_{-}(i \omega)][1-\tilde{\psi}_{+}(i \omega)]}{1-\tilde{\Psi}_{-}(i \omega)\tilde{\psi}_{+}(i \omega)}\right\},
\label{2eq5}
\end{eqnarray}
where $\tilde{\psi}_{+}(i \omega) $$= $$\exp{(-i \omega T_E)}$  is the Laplace transform of $\psi_{+}$,  $\tilde{\Psi}_{-}(i \omega)$ is the Laplace transforms of $\Psi_{-}$ from Eq.~(\ref{RTD1}), $\Delta x$$ =$$2$ and $\langle T_{\pm} \rangle$ are the mean residence times in the states $s$$=$$+1$ and $s$$=$$-1$ of the two-state process, respectively. It is clear that $\langle T_{+} \rangle$$ = $$T_E$ and $\langle T_{-} \rangle$$ = $$T_R + \int_{0}^{\infty} \xi \psi_{-}(\xi)\,d \xi$.

We compare the analytic power spectrum (\ref{2eq5}) with the numerical power spectrum of the original bistable system (\ref{1eq2}) calculated with the parameters $k $$=$$ -0.05$, 
$T$$ =$$ 50$, $D $$=$$ 0.053$. Note that any switching event in the bistable system has a finite duration, i.e. the change of the coordinate $x$ from $(+1)$ to  $(-1)$ takes certain time. This leads to the increase of the duration of locking. Therefore to successfully match the parameters of the bistable system (\ref{1eq2}) to the parameters of the two-state model we set $T_E$$=$$T_R$$=$$T$$ + $$\Delta$,  where $\Delta$ is the time needed for the particle to descend from the maximum of the effective potential $U_{\rm eff}$ into its minimum. $\Delta$ is approximately set to $5$ for $|k|$$<$$0.05$. 

\begin{figure}
\begin{center}
\includegraphics[width=0.45\textwidth]{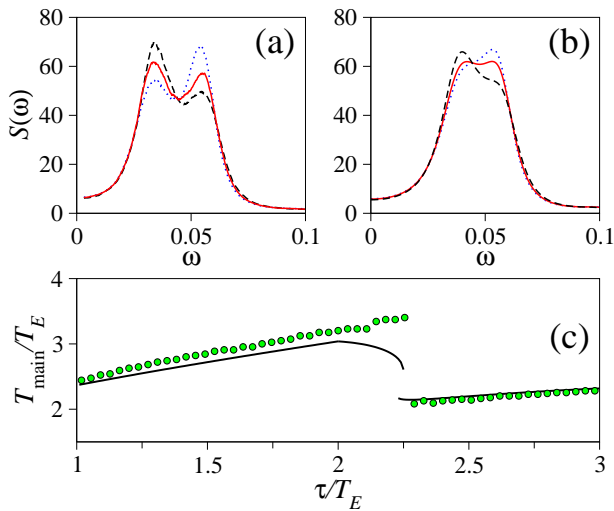}     
\caption{\label{fig3} (a) Numerical power spectrum of the bistable system (\ref{1eq2}) for $T = 50$ and three different $\tau$ close to $\tau_c$. (b) Analytic power spectrum (\ref{2eq5}) for the same parameters as in (a) and $T_E$$=$$T_R$$=55$. (c) Numerical (circles) and analytic (line) main period of the oscillations. }
\end{center}
\end{figure}

For the delay time $\tau$ between $2 T_E$ and $3 T_E$ the power spectrum has two maxima as shown in Fig.~\ref{fig3}(a). For $\tau$ close to $2 T_E$ the left maximum is higher than the right one (dashed line in Fig.~\ref{fig3}(a)). As $\tau$ increases from $2 T_E$ to $3 T_E$, the left maximum of $S(\omega)$ decreases and the right maximum increases. At critical $\tau_c$ the two maxima have equal heights as shown by the solid line in Fig.~\ref{fig3}(a). For $\tau$$>$$\tau_c$ the right maximum is higher than the left one as shown by the dotted line in Fig.~\ref{fig3}(a). Fig.~\ref{fig3}(b) shows the analytic power spectrum for the same parameters as in Fig.~\ref{fig3}(a).

The main period increases almost linearly with $\tau$ for $\tau$$<$$\tau_c$, at 
$\tau$$ =$$ \tau_c$ the main period drops discontinuously  and for $\tau$$ >$$ \tau_c$ it increases almost linearly again. This is shown in Fig.\,\ref{fig3}(c), where numerical results (circles) are compared with main period computed from the analytical power spectrum (\ref{2eq5}) (solid line). 

\subsection{Large delay times $\tau$: Equilibrium RTDs.}
\label{sec3}
\begin{figure}
\begin{center}
\includegraphics[width=0.45\textwidth]{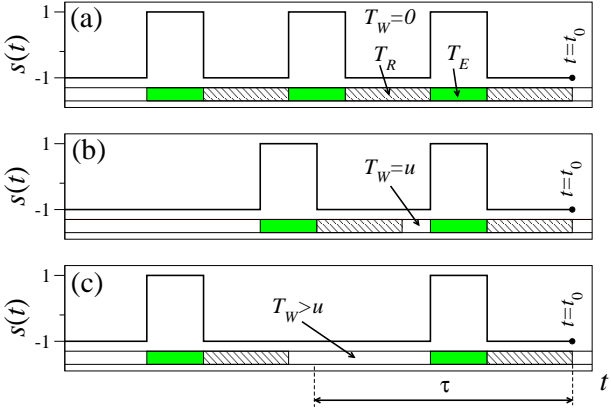} 
\caption{\label{fig4} Schematic representation of the configuration of pulses in the case of $\tau \in [T_E + 2 T_R; 2 T_E + 3 T_R]$. (a) The durations of the two latest waiting times are $T_W=0$. (b) Durations of the latest waiting time is $T_W=u>0$. (c) Durations of the latest waiting time is $T_W>u$.}
\end{center}
\end{figure}

Now consider the case of the delay times larger than $(T_E$$ +$$ 2 T_R)$. This situation is qualitatively different from the one discussed in Section \ref{sec2a} because for large delay times the history is variable. In other words, there are many possible profiles of the two-state system on the interval of time $[t_0-\tau;t_0]$, where $t_0$ is the time moment when the current refractory phase has just finished. This situation is sketched in Fig.~\ref{fig4}, where the time moment $t_0$ is indicated by the filled circle. The  distribution of the residence times in the state $s$$=$$-1$ changes in time and now the results of the renewal theory cannot be applied directly. To overcome this problem we introduce the concept of equilibrium RTDs in the sense of the averaging over all possible histories. The equilibrium RTDs can be computed from the known instantaneous RTDs as  shown below. 

Consider a certain history $\tilde{s}(t)$ of the two-state stochastic process $s(t) $$=$$ \pm 1$ on the interval of time $t \in [t_0-\tau;t_0]$ inside which there are $\tilde{N}_{\tau}$ pulses. Note, that inside an interval of length $\tau$ there can be not more than $N_{\tau}$ pulses, where $N_{\tau}$$=$$[\tau/(T_E+T_R)]$ and $[\cdots]$
denotes an integer part of the number. Number these pulses with index $j$ changing from $1$ to $\tilde{N}_{\tau} \le N_{\tau}$, with $j$$=$$1$ corresponding to the pulse closest to $t_0$. 
Denote by $u_j$ the duration of the waiting phase that precedes the $j$-th pulse.
If we know the  number $\tilde{N}_{\tau}$ of pulses inside $t \in [t_0-\tau;t_0]$ together with the ordered durations of the waiting phases
$u_j, \ j=1,2,\ldots,\tilde{N}_{\tau}$, we can unambiguously reconstruct $\tilde{s}(t)$. Let us consider all possible histories $\tilde{s}_k(t)$ and form an $N_{\tau}$-dimensional ``history'' space $H^{\tau}_{\mathbf u}$ with vectors ${\mathbf u}_k$$=$$(u_{k,1},u_{k,2},\ldots,u_{k,N_{\tau}})$. If the number of pulses $\tilde{N}_{\tau,k}$
for the given history $\tilde{s}_k(t)$ is less than the largest possible number $N_{\tau}$, the redundant coordinates $u_{k,j}, \ j$$=$$\tilde{N}_{\tau,k}$$+$$1,\ldots,N_{\tau}$ are set to zero. Namely, if there is only one pulse on the given interval, we set $u_{k,1}$$=$$(\tau$$-$$T_E$$-$$T_R)$ and $u_{k,j}$$=$$0, \  j$$=$$2,\ldots,N_{\tau}$. If there are two pulses, we set
$u_{k,2}$$=$$(\tau$$-$$2T_E$$-$$2T_R$$-$$u_{k,1})$ and $u_{k,j}$$=$$0, \  j$$=$$3,\ldots,N_{\tau}$, etc. Therefore, the durations $u_{k,j}$ cannot be larger than $(\tau-T_E-T_R)$ by the way of construction. 
Obviously,  $H^{\tau}_{\mathbf u}$ is in one-to-one correspondence with the
set of all possible histories $\tilde{s}_k(t)$. 

Next, split the space $H^{\tau}_{\mathbf u}$ into $M$ equal $N_\tau$-dimensional cells $I_l$,\,$l=$\,$1,2,...,M$ of the form $I_l$$=$$[u_{l,1} \pm \eta/2] \times [u_{l,2} \pm \eta/2] \times ... \times [u_{l,N_\tau} \pm \eta/2] $, where $\eta$ is the side length of each cell. 
$H^{\tau}_{\mathbf u}$ can be represented as a unity of $N_{\tau}$-dimensional cells as follows:
\[ H^{\tau}_{\mathbf u}=\big\{ I_1  \cup I_2 \cup ... \cup I_{M} \big\}.  \]
Consider one cell $I_k$. One can assume that all histories corresponding to the points inside this cell are approximately the same. Therefore, 
the waiting times that start immediately after $t_0$ and whose histories are approximately ${\mathbf u}_k$, have approximately the same 
RTDs $\psi_{-}^{\mathbf{u}_k}$ which are solutions of Eq.~(\ref{2eq1}).

Now consider some sufficiently long realization of the two-state stochastic process $s(t)$.
Suppose we are interested in some function $f$ of the randomly changing waiting time 
$\xi$, and we wish to compute an average of $f$.  There are two kinds of averages: over time and over the ensemble of realizations, which coincide if the underlying process is ergodic. The time average $\langle f \rangle$ can be calculated from a single realization as follows
\begin{eqnarray}
\langle f \rangle = \lim_{N \rightarrow \infty}\frac{1}{N} \sum_{i=1}^{N} f(\xi_i), 
\label{3eq1}
\end{eqnarray}
where $N$ stands for the number of pulses in the realization. In a sufficiently long realization one can find a large number $n_k$ of 
histories
from the cell $I_k$. The waiting times occurring after these histories have (approximately) the same distribution $\psi_{-}^{{\mathbf u}_k}$. In an infinitely long realization when $N$$ \rightarrow$$ \infty$, one can find {\it all} histories from the same cell, i.e. $n_k \rightarrow \infty$, with every history being found an infinite number of times. 
We now regroup the summands in  Eq.~(\ref{3eq1}) by collecting inside each bracket number $k$ the values of $f$ of the waiting times $\xi$ occurring after the history from the same cell centered at  ${\mathbf u}_k$. We denote the respective values of $f$ as $f_l^{{\mathbf u}_k}$, $l=1,\ldots,n_k$. Also, each bracket number $k$ is divided and multiplied by $n_k$. 
\begin{eqnarray}
\langle f \rangle &=& \lim_{N \rightarrow \infty} \bigg( \frac{n_1}{N}\frac{1}{n_1}\underbrace{[f_{1}^{{\mathbf u}_1}+f_{2}^{{\mathbf u}_1}+...+f_{n_1}^{{\mathbf u}_1}]}_{n_1}   \nonumber \\ 
&+& \frac{n_2}{N}\frac{1}{n_2}\underbrace{[f_{1}^{{\mathbf u}_2}+f_{2}^{{\mathbf u}_2}+...+f_{n_2}^{{\mathbf u}_2}]}_{n_2}  \nonumber \\ &+&  ...+ \frac{n_M}{N}\frac{1}{n_M}\underbrace{[f_{1}^{{\mathbf u}_M}+f_{2}^{{\mathbf u}_M}+...+f_{n_M}^{{\mathbf u}_M}]}_{n_M}\bigg)   ,
\label{3eq2}
\end{eqnarray}
%

Introduce the equilibrium RTD $\psi_{-}^{\rm eq} (\xi)$ in the sense that $\psi_{-}^{\rm eq} (\xi)\,d\xi$ gives the probability for the waiting time to  have the duration in the interval $[\xi;\xi+d\xi]$. The equilibrium RTD does not depend on history, i.e. $\psi_{-}^{\rm eq} (\xi)\,d\xi$ is understood as the number $N_\xi$ of the waiting times  with the duration $T_W$ that fall within the interval $[\xi;\xi+d\xi]$ divided by the total number $N$ of pulses in any given realization of the stochastic process, in the limit as $N \rightarrow \infty$
\begin{eqnarray}
\psi_{-}^{\rm eq} (\xi)\,d\xi = \lim_{N \rightarrow \infty} \frac{N_\xi}{N}.
\label{3eq3}
\end{eqnarray}
Clearly, the factors $n_k / N$ in (\ref{3eq2}) are the probabilities of the history to be in the
cell $I_k$. Denote these probabilities by ${\mathcal P}({\mathbf u}_k)\,(d\eta)^{N_\tau}$, where ${\mathcal P}({\mathbf u})$ is the corresponding distribution density. From the general considerations it is clear that ${\mathcal P}({\mathbf u})$ depends solely on $\psi_{-}^{\rm eq}$. The particular dependence of ${\mathcal P}({\mathbf u}_k)$ on $\psi_{-}^{\rm eq}$ is to be determined separately for any given history ${\mathbf u}_k$. 

The terms $\frac{1}{n_k}\underbrace{[f_{1}^{u_k}+f_{2}^{u_k}+...+f_{n_k}^{u_k}]}_{n_k} $ in (\ref{3eq2}) are the time averages of $f(\xi)$, for $\xi$ with the history determined by the cell centered around ${\mathbf u}_k$ and  with distribution density $\psi_{-}^{{\mathbf u}_k}$. Now, since the process is assumed to be ergodic, a time average is equal to the ensemble average. The latter can be calculated with the knowledge of the RTD $\psi_{-}^{{\mathbf u}_k}$ for the given history as 
$\int_{0}^{\infty} \psi_{-}^{u_k}(\xi) f(\xi)\, d\xi$.
  
Finally, in the limit $M \rightarrow \infty$ and $\eta \rightarrow 0$, the equation (\ref{3eq2}) can be rewritten in the form
\begin{eqnarray}
\langle f \rangle &=& \int_{0}^{\infty} \psi_{-}^{\rm eq} (\xi) f(\xi)\, d \xi \nonumber \\
&=& \int_{u \in H_{\mathbf u}^{\tau}} {\mathcal P} ({\mathbf u})\, du_1 du_2 ...du_{N_\tau} \int_{0}^{\infty}\psi_{-}^{{\mathbf u}} (\xi) f(\xi) \,d \xi. \nonumber \\
\label{3eq4}
\end{eqnarray}
Regroup terms in Eq.~(\ref{3eq4}) as follows
\[ 0= \int_{0}^{\infty}f(\xi) d \xi  
\underbrace{ \bigg\{- \psi_{-}^{\rm eq} (\xi) + \int_{u \in H_{\mathbf u}^{\tau}} 
  {\mathcal P} ({\mathbf u}) \psi_{-}^{{\mathbf u}} (\xi)\, du_1 du_2 ... du_{N_\tau} \bigg\}  }_{=0}   \]
Note that Eq.~(\ref{3eq4}) holds for an arbitrary integrable function $f$, which means that 
the expression in brackets is equal to zero. 
We therefore arrive at the following integral equation for the unknown equilibrium RTD
\begin{eqnarray}
\psi_{-}^{\rm eq} (\xi) = \int_{{\mathbf u} \in H_{\mathbf u}^{\tau}} {\mathcal P} ({\mathbf u}) \psi_{-}^{\mathbf u} (\xi)\,du_1 du_2 ... du_{N_\tau}.
\label{3eq5}
\end{eqnarray}

Equation (\ref{3eq5}) allows us to treat the renewal processes with non-identically distributed waiting times. It generalizes the results of the renewal theory \cite{Cox}.  

\subsection{Equilibrium RTD for $\tau \in [T_E + 2 T_R; 2 T_E + 2 T_R]$}
\label{sec3a}
We now use the derived equation (\ref{3eq5}) to calculate the equilibrium RTD $\psi_{-}^{\rm eq}$ in the case of $\tau \in [T_E + 2 T_R; 2 T_E + 2 T_R]$.
The reason for choosing this interval for $\tau$ instead of the whole interval $[T_E + 2 T_R; 2 T_E + 3 T_R]$, where the history space 
$H_{\mathbf u}^{\tau}$
is one-dimensional is the following. If $\tau \in [T_E + 2 T_R; 2 T_E + 2 T_R]$ the history can contain only one complete pulse and one half-complete pulse (see Fig.~\ref{fig4}), depending on the duration of the previous waiting phase $T_W$. Therefore, this situation is simpler than the case of $\tau \in [T_E + 2 T_R; 2 T_E + 3 T_R]$, where the history can contain up to two complete pulses.
 Denote the duration of the previous waiting phase by $u$ and consider two possible cases.

{\it Case 1:} For $u \in [ \tau -T_E -2 T_R;\infty)$ the history contains one complete pulse. In this case the solution of Eq.~(\ref{2eq1}) is given by
\begin{eqnarray}
P_{-}^{0}(\xi) &=& e^{-p\xi} \\
&\times&
\left\{
\begin{array}{lcl}
1,&\ &\xi\in [0;\tau - T_E - T_R] \\
e^{-q \xi + q (\tau - T_E -T_R )},&\ &
\xi \in [\tau - T_E -T_R;\\
&\ & \qquad \tau - T_R] \\
e^{-q T_E},&\ &\xi \in [\tau -  T_R;\infty).
\end{array}
\right.\nonumber 
\label{3aeq1}
\end{eqnarray}
{\it Case 2:} For $u \in [0; \tau -T_E -2 T_R]$ the history contains one complete pulse and one half-complete pulse. In this case the solution of Eq.~(\ref{2eq1}) is
\begin{eqnarray}
P_{-}^{u}(\xi) &=& e^{-p \xi}\\
&\times&\left\{
\begin{array}{lcl}
e^{-q \xi },&\ &\xi\in [0;\\
&\ & \qquad \tau - T_E -2 T_R -u] \\
e^{-q (\tau - T_E -2 T_R -u)},&\ &\xi \in [\tau - T_E -2 T_R -u;\\
&\ & \qquad \tau -T_E -T_R] \\
e^{-q \xi + q (T_R + u)},&\ &\xi \in [\tau - T_E - T_R;\\
&\ & \qquad \tau - T_R]\\
e^{-q (\tau - 2 T_R -u)},&\ &\xi \in [\tau -T_R;\infty).
\end{array}
\right. \nonumber 
\label{3aeq2}
\end{eqnarray}
The connection between the probability density ${\mathcal P}(u)$ and $\psi_{-}^{\rm eq}$ is straightforward
\begin{eqnarray}
{\mathcal P}(u) = \psi_{-}^{\rm eq}(u).
\label{3aeq3}
\end{eqnarray}
Because $u$ is constrained to the interval $[0;\tau - T_E -2 T_R]$, the relation Eq.~(\ref{3eq5}) is an equation for the unknown $\psi_{-}^{\rm eq}(\xi)$ only on the same interval of $\xi \in [0;\tau - T_E -2 T_R]$. For $\xi>(\tau - T_E -2 T_R)$, Eq.~(\ref{3eq5}) is no longer an equation, as will be shown below. 

For $\xi \in [0;\tau - T_E -2 T_R]$ we divide the integration interval $U$ according to the restrictions on $\xi$ and $u$ as in (\ref{3aeq2}) and rewrite Eq.~(\ref{3eq5}) as follows
\begin{eqnarray}
\psi_{-}^{\rm eq}(\xi) = p e^{-p \xi} \left( 1- \int_{0}^{\tau - T_E -2 T_R} \psi_{-}^{\rm eq} (u)\,d u \right)  \label{3aeq4} \\
+(p +q)e^{-(p + q)\xi}\int_{0}^{\tau - T_E - 2 T_R - \xi}\psi_{-}^{\rm eq} (u)\,d u \nonumber \\ 
+ p e^{-p \xi}e^{-q (\tau - T_E - 2 T_R)}\int_{\tau - T_E -2 T_R - \xi}^{\tau - T_E -2 T_R }e^{q u}\psi_{-}^{\rm eq} (u)\,d u.\nonumber  
\end{eqnarray}
We look for the solution of Eq.~(\ref{3aeq4}) in the form 
\begin{eqnarray}
\psi_{-}^{\rm eq}(t) &=& A e^{-(p +q) \xi}, 
\label{3aeq5}
\end{eqnarray}
where $A$ is some unknown constant.
Plugging (\ref{3aeq5}) into (\ref{3aeq4}) and comparing the coefficients of the exponents $e^{-p \xi}$ and  $e^{-(p + q)  \xi}$ yields the constant $A$ 
\begin{eqnarray}
A = \frac{p(p+q)}{p + q\,e^{-(p +q)(\tau - T_E - 2 T_R)}}.
\label{3aeq6}
\end{eqnarray}
On the interval $\xi \in [\tau- T_E -2 T_R;\tau - T_E - T_R]$  Eq.~(\ref{3eq5}) becomes
\begin{eqnarray}
\psi_{-}^{\rm eq}(\xi) = p e^{-p \xi} \left( 1- \int_{0}^{\tau - T_E -2 T_R} \psi_{-}^{\rm eq} (u)\,d u \right)  \nonumber \\
+ p e^{-p \xi}e^{-q (\tau - T_E - 2 T_R)}\int_{0}^{\tau - T_E -2 T_R }e^{q u}\psi_{-}^{\rm eq} (u)\,d u,\label{part2}
\end{eqnarray}
which is no longer an equation, because the r.h.s. of Eq.~(\ref{part2}) depends on the known $\psi_{-}^{\rm eq}$ on the interval $\xi \in [0;\tau - T_E - 2 T_R]$. 
Similarly, the equilibrium RTD $\psi_{-}^{\rm eq}$ on the interval $\xi \in [\tau- T_E - T_R;\tau - T_R]$ is determined as 
\begin{eqnarray}
\psi_{-}^{\rm eq}(\xi) &=& (p+q) e^{-(p+q) \xi} e^{q(\tau-T_E -T_R)}   \label{part3} \\
&\times&\left( 1- \int_{0}^{\tau - T_E -2 T_R} \psi_{-}^{\rm eq} (u)\,d u \right)  \nonumber \\
&+& (p+q) e^{-(p+q) \xi}e^{q T_R}\int_{0}^{\tau - T_E -2 T_R }e^{q u}\psi_{-}^{\rm eq} (u)\,d u.\nonumber 
\end{eqnarray}
Finally,  the equilibrium RTD $\psi_{-}^{\rm eq}$ on the interval $\xi \in [\tau- T_R;\infty)$ reads
\begin{eqnarray}
\psi_{-}^{\rm eq}(\xi) &=& p e^{-p \xi} e^{-q T_E}\left( 1- \int_{0}^{\tau - T_E -2 T_R} \psi_{-}^{\rm eq} (u)\,d u \right)  \nonumber \\
&+& p e^{-p \xi}e^{-q (\tau - 2 T_R)}\int_{0}^{\tau - T_E -2 T_R }e^{q u}\psi_{-}^{\rm eq} (u)\,d u.\nonumber \\  
\label{part4}
\end{eqnarray}
Taking the integrals in Eqs.~(\ref{part2})--(\ref{part4}) with $\psi_{-}^{\rm eq}$ on the interval $\xi \in [0;\tau - T_E - 2 T_R]$ from Eq.~(\ref{3aeq5}), one obtains the equilibrium RTD on the whole interval 
\begin{eqnarray}
&&\psi_{-}^{\rm eq}(\xi) = A  \label{3aeq7} \\
&\times& \left\{
\begin{array}{lcl}
 e^{-(p+q) \xi},&\ & \xi\in [0;\tau - T_E-2 T_R] \\
 e^{-q (\tau - T_E - 2 T_R)-p \xi},&\ & 
 \xi \in [\tau - T_E - 2T_R; \\
 & \ & \qquad \tau - T_E-T_R] \\
\frac{(p + q)}{p} e^{-(p+q) \xi}e^{q T_R},&\ & 
 \xi \in \ [\tau -  T_E - T_R;\tau -T_R]\\
e^{-q (\tau-2 T_R)-p \xi},&\ & \xi \in [\tau -  T_R;\infty).
\end{array}
\right. \nonumber
\end{eqnarray}
The knowledge of the RTD allows us to compute the average of any given function  $g(s)$ of the stochastic variable $s(t)$
\begin{eqnarray}
\langle g \rangle = \frac{g(-1) \langle T_{-}\rangle + g(+1)T_E}{T_E + \langle T_{-}\rangle  },
\label{3aeq8}
\end{eqnarray}
where $\langle T_{-}\rangle = T_R + \int_{0}^{\infty} u \psi_{-}^{\rm eq} (u)\, d u$ for $\tau \in [T_E + 2 T_R; 2 T_E + 2 T_R]$. For the values of the delay time $\tau < T_E + 2 T_R$ the equilibrium RTD must be replaced by the corresponding $\psi_{-}(u)$ from Eq.~(\ref{2eq1}) with $P_{-}(u)$ from Eqs.~(\ref{2eq3}) or (\ref{2eq4}). 

\begin{figure}
\begin{center}
\includegraphics[width=0.45\textwidth]{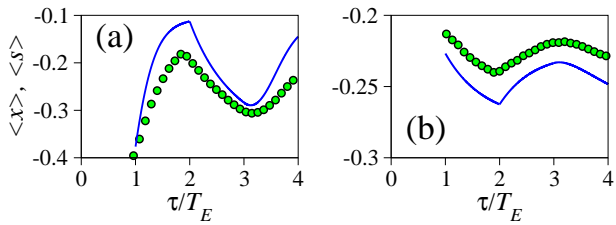}
\caption{\label{fig5} Comparison of the average $x$ {\it vs} delay time calculated numerically (circles) with the average $\langle s \rangle$ of the two-state process (\ref{3aeq8}) (lines). Parameters are (a): $D = 0.043$, $T = 50$, $T_E = T_R =55$, $k=0.05$ and (b): $D = 0.053$, $T = 50$, $T_E = T_R =55$, $k=-0.01$. }
\end{center}
\end{figure}

We present here the results of the comparison of the analytic formula Eq.~(\ref{3aeq8}) with the simulation of the original bistable system (\ref{1eq2}). Figs.~\ref{fig5}(a)-(b)  show the average $\langle x \rangle$ {\it vs} delay time $\tau$ for $D=0.043$, $k=0.05$ and  $D=0.053$, $k=-0.01$, respectively.   Analytic result (\ref{3aeq8}) (lines) is compared with $\langle x \rangle$ calculated numerically (circles). As in Fig.~\ref{fig3} the durations of the excited and the refractory phases are $T_E = T_R = 55$. 

For positive feedback strength (Fig.~\ref{fig5}(a)) the average $\langle x \rangle$ increases with the delay until $\tau = 2 T_E$ then it decreases when $\tau$ is in the interval $\tau \in [2 T_E; 3 T_E]$ and after that it increases again when $\tau \in [3 T_E ; 4 T_E]$. The behaviour of $\langle x \rangle$ on $\tau$ is exactly the opposite for negative feedback strength $k=-0.01$  as it is shown in Fig.~\ref{fig5}(b). We do not plot the variance $\langle x^2 \rangle$ because it is independent of delay: according to (\ref{3aeq8}), for the stochastic process $s= \pm1$ the variance is constant $(\langle s^2 \rangle = 1)$. 

Note that the non-monotonous dependence of the average $\langle x \rangle$ on the delay time is qualitatively similar to the dependence of the variance $\langle x^2 \rangle$ on delay time which was derived in \cite{PJ07a} for the van der Pol oscillator near the Andronov-Hopf bifurcation. This similarity shows again that the properties of the noise-induced oscillations in the excitable systems are closely related to the properties of non-excitable noise-driven systems near bifurcations \cite{Janson04}.
\section{Power spectrum in the mean field approximation}
\label{sec4} 
The knowledge of the equilibrium RTD does not allow us to compute the power spectrum of noise-induced oscillations. According to \cite{Mel93} the power spectrum is given by the Fourier transform of the average product of probability currents $\langle j(t) j(t^\prime)\rangle$, where the current $j(t)$ is a sum of $\delta$-like pulses occurring at the moments $t_n$ of switching: $j = 2 \sum_{n} (-1)^n \delta \left( t - t_n\right)$ . Any switching time $t_n$ is the time moment, when the transition from the state $s$\,$=$\,$-1$ to the state $s$\,$=$\,$+1$ or backwards occurs. The amplitude of the current $j$ is given by $+2$ in case when $s$\,$=$\,$-1$ changes to $s$\,$=$\,$+1$ and by $-2$ otherwise.  Unfortunately, the average $\langle j(t) j(t^\prime)\rangle$ can not be represented in the form (\ref{3eq2}) and therefore the correlation function can not be written in terms of the equilibrium $\psi_{-}^{\rm eq}$. 

One possible approximation for the power spectrum 
is an analog of the mean field approximation, when the known result for the power spectrum from the renewal theory \cite{Cox,Goy04} is calculated with the equilibrium $\psi_{-}^{\rm eq}$ obtained above. 

\begin{figure}
\begin{center}
\includegraphics[width=0.45\textwidth]{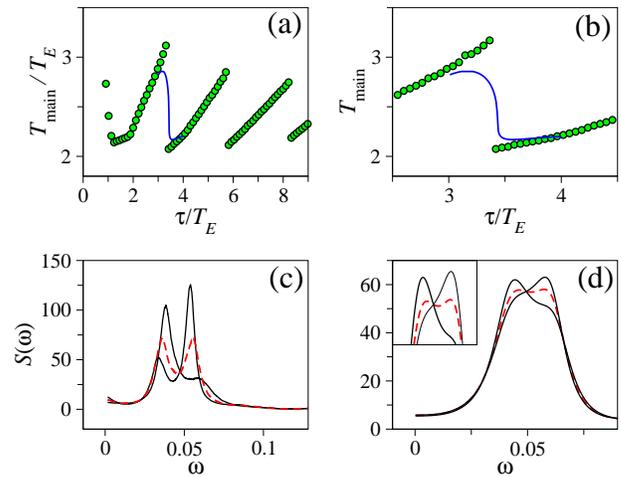}
\caption{\label{fig6} (a)  Circles represent the main period $T_{\rm main}$ of the oscillations in units of $T_E = T_R =55$  calculated numerically, solid line shows the corresponding analytic counterpart computed from (\ref{2eq5}) and (\ref{3aeq7}). (b) Zoom of the region in (a) near $\tau = 3 T_E$. (c) Numerical power spectrum for $\tau$ near the critical $\tau_c$. Dashed line shows power spectrum at $\tau = \tau_c$. (d) Analytic power spectrum for $T_E = T_R = 49$. The inset in (d) shows zoom of the area around the maximum of $S(\omega)$.  }
\end{center}
\end{figure}

We compare this version of the mean field approach with the simulation of the bistable system (\ref{1eq2}) in Fig.~\ref{fig6}. For $D = 0.043$, $k=0.05$ and the rest of the parameters as in Fig.~\ref{fig5}, in Figs.~\ref{fig6}(a,b) the main period $T_{\rm main}$ of the noise-induced oscillations  is given as a function of  delay time $\tau$. Both $T_{\rm main}$ and $\tau$ are shown in the units of $T_E = T_R$. Circles in Figs.~\ref{fig6}(a,b) correspond to the results of simulation, lines show the main period computed from the analytic expression for the power spectrum (\ref{2eq5}) with the equilibrium RTD (\ref{3aeq7}). For simplicity we show only the part of the analytical curve for $\tau \in [3 T_E; 4 T_E]$. As we see, the mean-field analytics predicts correctly the critical delay time where the first branch-switching occurs. 

In Fig.~\ref{fig6}(c) the numerical power spectrum is shown for three different values of $\tau$ close to its critical value $\tau_c$.  Fig.~\ref{fig6}(d) shows the analytic power spectrum for the same parameters as in  Fig.~\ref{fig6}(c) and $T_E = T_R = 49$. 
\section{Incoherence maximization due to delay}
\subsection{Bistable excitable system}
\label{sec5} 
The RTDs $\psi_{-}$ 
derived in  Section~\ref{sec2}, along with the equilibrium RTD (\ref{3aeq7}), allow us to compute the coefficient of variation $R$ \cite{Lind04}, which serves as a measure of coherence of spiking 
\begin{eqnarray}
\label{coher}
R = \frac{\sqrt{\langle T_{-}^2 \rangle - \langle T_{-} \rangle ^2}}{T_E + \langle T_{-} \rangle}.
\end{eqnarray}
The smaller is the coefficient of variation $R$ the more regular is the spiking. 
\begin{figure}
\begin{center}
\includegraphics[width=0.45\textwidth]{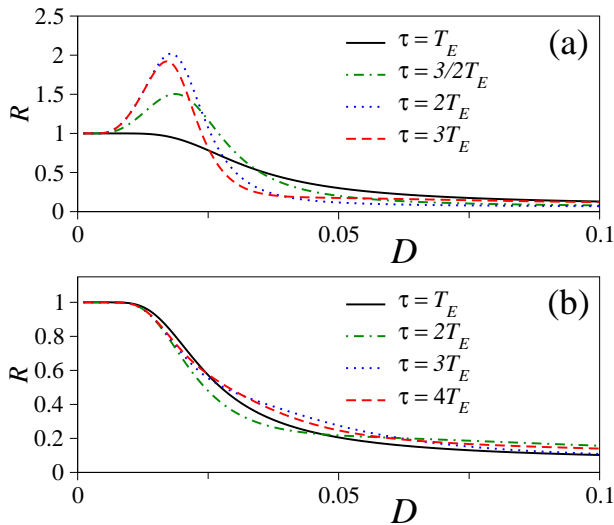}
\caption{\label{fig7} 
(a) Coefficient of variation $R$ {\it vs} noise strength $D$ for $k=0.05$ and $T_E = T_R =55$. (b) The same as in (a) for the negative feedback strength $k=-0.05$.
}
\end{center}
\end{figure}
The effect of coherence resonance \cite{Pik97} manifests itself in the appearence of a minimum in the dependence of the coefficient of variation  on the noise strength $D$. The opposite effect to coherence resonance is regarded as incoherence maximization \cite{Lind04}. It appears when there is a local maximum in the dependence of $R$ on $D$.
 
Here we show that in the bistable system presented in the Section\,\ref{sec1}, delay induced incoherence maximization can be observed for positive feedback strength $k$. In terms of the two-state model with history dependent transition rates Tab.\,(\ref{tab1}), positive feedback strength $k$ corresponds to the enequality $p+q > p$, implying that the probability of transition from the state $s$\,$=$\,$-1$ to the state $s$\,$=$\,$+1$ is larger if $s_{\tau}$\,$=$\,$+1$. 
Since the durations of the excited $T_E$ and the refractory $T_R$ states are fixed, the dependence on noise is realized through the duration of the waiting phase $T_W$ only. Obviously, in the limit of large noise strength ($D$\,$\rightarrow$\,$\infty$), the duration $T_W$ becomes vanishingly small, leading to almost regular spiking with the mean interspike interval given by $T_E + T_R$. For vanishing noise ($D$\,$=$\,$0$) the transition probability is neglidgibly small resulting again in the regular spike train. Therefore, for some finite noise strength we can expect to observe a local maximum of the spike incoherence.

Using the Kramers relation (\ref{1eq5}) \cite{Kram,Gard} between the transition rates $p$ and $q$ and the noise strength $D$, we plot in Fig.\,(\ref{fig7}) the coefficient of variation $R$ as function of $D$. In Fig.~\ref{fig7}(a) $R$ is shown for positive feedback strength $k=0.05$, $T_E = T_R =55$ and different delay times $\tau$ as indicated in the legend. With no feedback $(\tau \le T_E)$ the coefficient of variation decreases monotonically with $D$. However, at any $\tau$ larger than, and close to, $T_E$ a maximum appears at a certain noise strength $D$. This is an evidence of incoherence maximization induced by the delay.  
If the feedback strength is negative $k=-0.05$ (Fig.~\ref{fig7}(b)) the incoherence maximization is absent at least for the delay times $\tau \le 4 T_E$. 

\subsection{Comparison with the FitzHugh-Nagumo system}
\label{sec5a}
To demonstrate incoherence maximization predicted in the Section\,(\ref{sec5}), we use the FitzHugh-Nagumo system with non-linear delayed feedback introduced through the activator $x$ into the equation for inhibitor $y$
\begin{eqnarray}
\epsilon \dot{x} &=& x-\frac{x^3}{3}-y, \nonumber \\
\dot{y} &=& x+a - K [x_{\tau}-x]^2 + D \zeta(t),
\label{FHN}
\end{eqnarray}
where $x_{\tau} = x(t-\tau)$, $\epsilon = 0.01$ and $a=1.1$.

The form of the feedback term in Eqs.\,(\ref{FHN}) was chosen in such a way, that the noise-induced dynamics in the waiting phase is similar to that of the two-state model introduced in the Section\,(\ref{sec2}). 

It should be emphasized that the FitzHugh-Nagumo system with {\it linear} feedback cannot be approximated by a two-state model with the transition rates as in Table\,(\ref{tab1}).
To see why this is so, consider a typical pulse train given by variable $x$ as function of time. This is shown in Fig.\,\ref{fig2a}(a). 
Assume that the system is currently in the waiting phase, $x(t) \in {\rm waiting}\,\,{\rm phase}$,  and assume further that $\tau$ seconds ago the system was in the refractory phase, $x(t-\tau) \in {\rm refractory}\,\,{\rm phase}$. 
Since the value of the state variable in the waiting phase is different from that in the refractory phase, unlike in the two-state model, we conclude that the feedback term which is given by $k  [x(t) - x(t-\tau)]$ is not zero.
However, this term becomes zero (up to fluctuations whose order is given by the noise strength $D$) if $\tau$ seconds ago the system was in the waiting phase.

This means that the current transition rate $\lambda$ from the waiting phase to the excited phase changes depending on whether $\tau$ seconds ago the system was in the waiting or in the refractory phase.
This contradicts the assumption that the transition rate depends on $\tau$ according to Table.\,(\ref{tab1}). 

The comparison between the FitzHugh-Nagumo model and the two-state model is possible only in the case of nonlinear feedback, e.g. like in Eqs.\,(\ref{FHN}) and strong time scale separation, i.e. $\epsilon$\,$\ll$\,$1$, when the concept of the transition rates can be applied.
The nonlinear feedback term ensures that during the waiting phase ($x(t) \approx -1.1$) the value of $[x_{\tau}-x]^2$ becomes significant only if $\tau$ second ago the system was in the excited phase, i.e. if $x_{\tau} \approx 2$. On the other hand, if $x_{\tau}$ belongs to the refractory phase, the term $[x_{\tau}-x]^2$ is negligibly small, so that the transition rate from the non-excited to the excited state is the same as in the original system without the feedback. Consequently, the transition rate is modified by the feedback only if $x_{\tau}$ belongs to the excited state. 

It is easy to see from Eqs.\,(\ref{FHN}) that negative values of the feedback strength $K$ effectively decrease the transition rate, whereas the positive values of $K$ increase the probability of transition. Therefore, based on the predictions made in the Section\,(\ref{sec5}), we conclude that delay-induced incoherence maximization should be observed for positive $K$.  
 
\begin{figure}
\begin{center}
\includegraphics[width=0.5\textwidth]{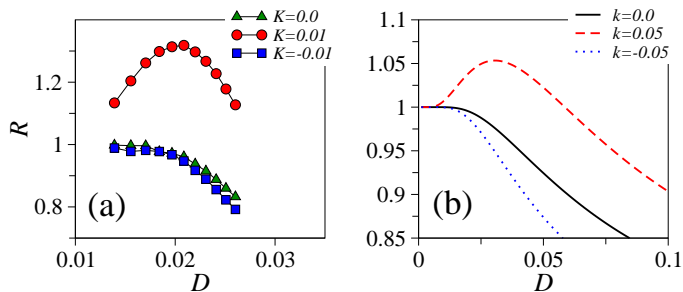}
\caption{\label{fig8}  
(a) Coefficient of variation $R$ {\it vs} noise strength $D$ for the FitzHugh-Nagumo system Eqs.\,(\ref{FHN}) with fixed delay time $\tau = 4$ and the feedback strength $K$ as indicated in the legend. (b) Coefficient of variation $R$ {\it vs} noise strength $D$ for the two-state model with the parameters $T_R = 2.5$, $\tau = 4$, and different feedback strength $k$ given in the legend.
}
\end{center}
\end{figure}

This result is confirmed by numerically computing the coefficient of variation $R$ {\it vs} noise strength $D$ for the FitzHugh-Nagumo system Eqs.\,(\ref{FHN}) at fixed delay time $\tau=4$, as shown in Fig.\,(\ref{fig8})(a).  For negative $K$ and $K=0$, $R$ decreases monotonically with $D$, however for positive $K$ the coefficient of variation reaches a local maximum confirming  incoherence maximization at certain noise strength $D$. 

To match the parameters of the two-state model with those of the FitzHugh-Nagumo system Eqs.\,(\ref{FHN}), we set $T_R = 2.5$, $T_E = 0.5$, $\tau=4$ and plot $R$ {\it vs} noise strength $D$ in Fig.\,(\ref{fig8})(b) at different values of the feedback strength $k$ (see legend). We see that the behaviour of the coefficient of variation $R$ on $D$ in the two-state model qualitatively coincides with the numerical results obtained for the FitzHugh-Nagumo system Fig.\,(\ref{fig8})(a). 
%
%

%
\section{Conclusion}
\label{concl}
To conclude, we presented a two-state model of an excitable system with time-delayed feedback. In this model the state variable $s$ takes only two values $s = \pm 1$ and the transition probability from one state into the other depends on the history of the process in a given way.  To compare the results derived for the two-state model with the properties of a real excitable system we consider a  bistable system (\ref{1eq2}) with the effective potential (\ref{1eq4}) which contains two delay times.  One of them is fixed and is used to model the excitability and the other one is assigned to the delay time of the controlling feedback force. 

Assuming that the durations of the excited phase and the refractory phase are noise-independent, we conclude that the only unknown and in general history-dependent quantity is the residence time density (RTD) $\psi_{-}(\xi)$ of the waiting phase. We show that for the delay times less than the sum of the duration of the excited phase $T_E$ and two durations of the refractory phase $2 T_R$, the history is non-variable, i.e. there is only one possible profile of the two-state process on the interval of time $[t_0-\tau;t_0]$, where $t_0$ is the time moment when the latest refractory phase has just finished. Therefore, all the waiting times are identically distributed and the renewal theory can be applied. In this case $\psi_{-}(\xi)$ as function of the delay time $\tau$ is computed straight forwardly  and represented by Eq.~(\ref{2eq1}) with $P_{-}$ determined by Eqs.~(\ref{2eq3}),(\ref{2eq4}). We use the results of the renewal theory \cite{Cox,Goy04} to obtain the power spectrum of the stochastic process $s(t)$. The main period of the noise-induced oscillations calculated from the analytically known power spectrum shows piece-wise linear dependence on the delay time. This analytical result confirms the similar finding obtained numerically in \cite{Janson04,Balanov04} for the FitzHugh-Nagumo system in the excitable regime.   

For the delay times larger than $(T_E + 2 T_R)$ the history becomes variable and the  distribution density of the waiting times is no longer time independent. To handle renewal processes with history-dependent RTDs the equilibrium RTD $\psi_{-}^{\rm eq}(\xi)$ Eq.~(\ref{3eq3}) in the sense of the averaging over all possible histories is introduced.
The integral Eq.~(\ref{3eq5}) for $\psi_{-}^{\rm eq}(\xi)$  is derived for an arbitrary delay time $\tau$. 
The solution of this equation is given for  $\tau \in [T_E + 2 T_R; 2 T_E + 2 T_R]$ by Eq.~(\ref{3aeq7}). The knowledge of $\psi_{-}^{\rm eq}(\xi)$ allows us to calculate the average of any given function $f(s)$ of the stochastic process $s$ be means of Eq.~(\ref{3aeq8}). This is an exact formula which is valid for any renewal process with the dependence on history.
Unfortunately, the results of the renewal theory can not be used directly to compute e.g. the power spectrum of the noise-induced oscillations.
 However, an analog of the mean-field approximation  is introduced when the expression for the power spectrum from the renewal theory is calculated with the equilibrium $\psi_{-}^{\rm eq}(\xi)$. 

The analytic results were compared with the results of numerical simulation of the bistable system (\ref{1eq2}). It is shown that the mean-field power spectrum predicts correctly the critical delay time when the first branch switching occurs in  the dependence of the main period {\it vs} delay time (Fig.~\ref{fig6}(a),(b)). 

Finally, we demonstrated incoherence maximization for positive feedback strength due to delay (Fig.~\ref{fig7}(a)). Hereby, the degree of incoherence measured by the coefficient of variation (\ref{coher}) is shown to posses a local maximum for increasing noise strength $D$ and fixed delay time. 
For negative feedback strength incoherence maximization is not observed up to the delay times of $(2T_E + 2 T_R)$.

\section{Acknowledgements}

The authors are grateful to A. Balanov who has made a number of helpful comments about the manuscript. This work was supported by EPSRC (UK). 


\end{document}